\documentclass[conference]{IEEEtran}
\IEEEoverridecommandlockouts
\usepackage{amsmath,amssymb,amsfonts}
\usepackage{algorithmic}
\usepackage{graphicx}
\usepackage{textcomp}
\usepackage{xcolor}
\usepackage{multirow}
\usepackage[utf8]{inputenc}
\usepackage{kotex}
\usepackage{caption}
\usepackage{graphicx}
\usepackage{amsmath}
\usepackage{booktabs}
\usepackage{multirow}
\usepackage{makecell}
\usepackage{rotating}
\usepackage{hyperref}

\usepackage[backend=biber,style=ieee]{biblatex} % Using IEEE style with biblatex
\def\BibTeX{{\rm B\kern-.05em{\sc i\kern-.025em b}\kern-.08em
    T\kern-.1667em\lower.7ex\hbox{E}\kern-.125emX}}
\begin{document}

\title{PixleepFlow: A Pixel-Based Lifelog Framework for Predicting Sleep Quality and Stress Level}

\author{\IEEEauthorblockN{Younghoon Na\textsuperscript{\dag}}
\IEEEauthorblockA{\textit{Seoul National University} \\
yh0728@snu.ac.kr}
\and
\IEEEauthorblockN{Seunghun Oh\textsuperscript{\dag}}
\IEEEauthorblockA{\textit{Hallym University}\\
gnsgus190@gmail.com}
\and
\IEEEauthorblockN{Seongji Ko\textsuperscript{\dag}}
\IEEEauthorblockA{\textit{Enssel Co., Ltd.}\\
sjko@gmail.com}
\and
\IEEEauthorblockN{Hyunkyung Lee\textsuperscript{*}}
\IEEEauthorblockA{\textit{Seoul National University} \\
hyunkyung913@snu.ac.kr}
\and

\thanks{\textsuperscript{\dag} These authors contributed equally to this work.}
\thanks{\textsuperscript{*} Corresponding author.}

}

\maketitle

\begin{abstract}
The analysis of lifelogs can yield valuable insights
into an individual's daily life, particularly with regard to their health and well-being.
The accurate assessment of quality of life is necessitated by the use of diverse sensors and precise synchronization.
To rectify this issue, this study proposes the image-based sleep quality and stress level estimation flow (PixleepFlow). PixleepFlow employs a conversion methodology into composite image data to examine sleep patterns and their impact on overall health. Experiments were conducted using lifelog datasets to ascertain the optimal combination of data formats. In addition, we identified which sensor information has the greatest influence on the quality of life through Explainable Artificial Intelligence(XAI). As a result, PixleepFlow produced more significant results than various data formats. This study was part of a written-based competition, and the additional findings from the lifelog dataset are detailed in Section \ref{Section:Extended Comparative Analysis}. More information about PixleepFlow can be found at\href{https://github.com/seongjiko/Pixleep}{https://github.com/seongjiko/Pixleep}. 

\end{abstract}

\begin{IEEEkeywords}
Lifelog, Sleep, Sleep Quality, Stress Level, Emotion Level, Deep Learning, Explainable AI, Wearable Devices, Health Monitoring, Image-Based Sleep Research, Multi-View Learning
\end{IEEEkeywords}

\section{Introduction}
Lifelogs, such as activity, heart rate and sleep data, reflect the quality of life. Changes in vital signals during daily activities and sleep can provide insights into sleep quality, emotions, and stress levels, which may have implications for overall health and well-being \cite{scataglini2023wearable,zhou2022population}.

Advanced sensors and deep learning models facilitate the inference of overall physical health from lifelog data without hospital visits. For instance, TzuAn Song's automated mobile sleep staging model (SLAMSS) predicts deep sleep (N3) with 79\% accuracy using wrist-worn actigraphy data \cite{song2023ai}. Rohit Gupta predicts stress and emotional states from wrist and chest sensor signals, achieving 95.54\% accuracy \cite{gupta2023multimodal}.
Hang Yuan developed a self-supervised deep learning model for sleep stage classification using wrist-worn accelerometers, a three-class (Wake/REM/NREM) classification F1 score of 0.57, indicating that the difference between polysomnography and model classification in external validation was a fairly accurate prediction of 34.7 minutes of total sleep time \cite{yuan2024self}. 
These advancements demonstrate the potential of data from smartphones and wearable to provide valuable insights into daily experiences and health monitoring.

\begin{figure}[ht]
\centering
\includegraphics[width=0.5\textwidth]{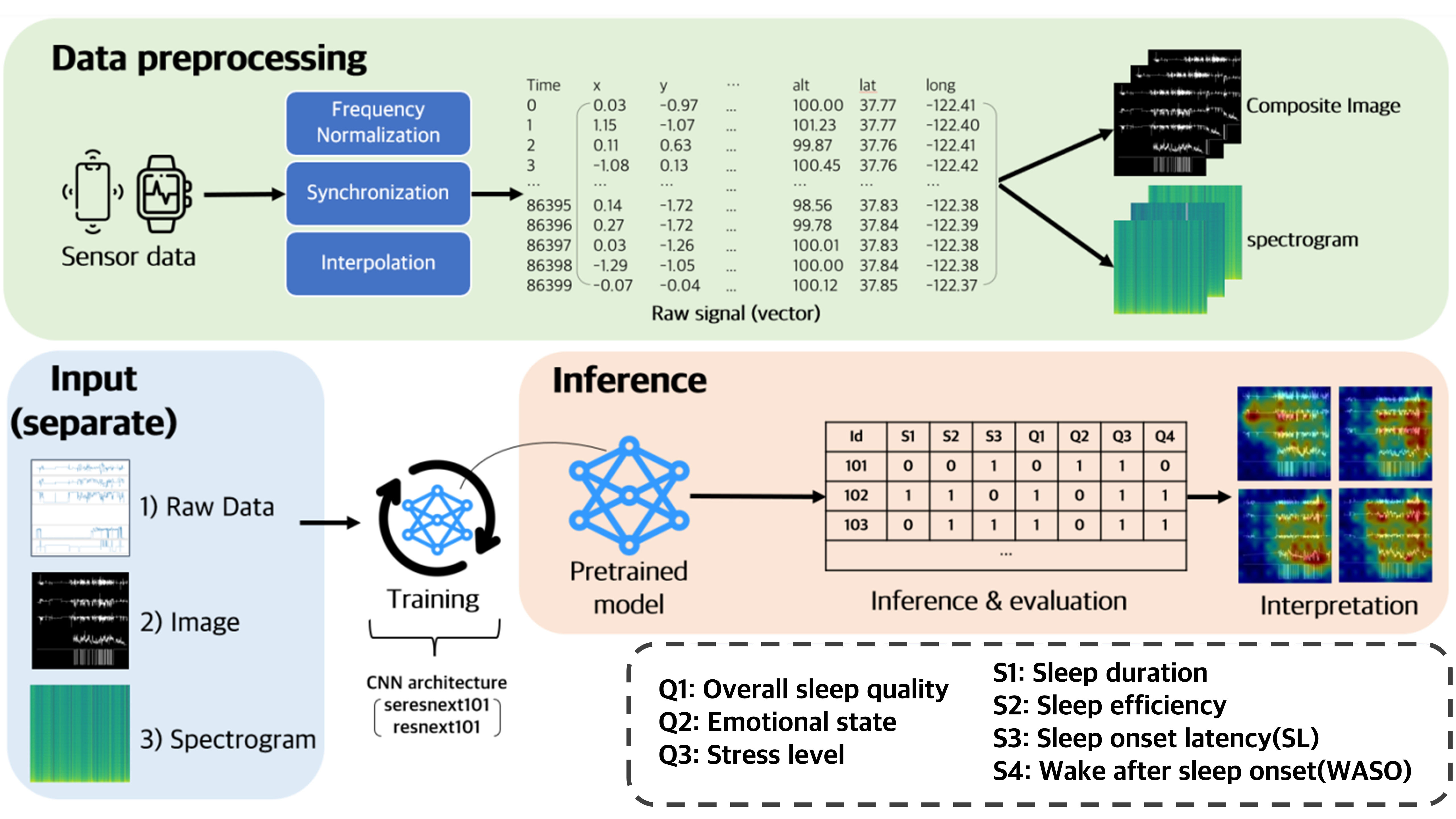}
\caption{PixleepFlow framework}
\label{fig:PixleepFlow}
\end{figure}

In most existing studies, sleep quality and stress level are predicted separately to assess the quality of life. However, we propose a more comprehensive approach by applying multi-label classification that simultaneously considers both sleep quality and stress level. To achieve this, we developed image-based sleep quality and stress level estimation flow (PixleepFlow), a model that leverages image-based input from various sensor data to concurrently predict seven key metrics related to sleep quality and stress level with high accuracy.

A key characteristic of PixleepFlow is its implementation of image-based inputs. By converting long-duration sensor data into image-based images, PixleepFlow effectively reduces high-dimensional data into three-dimensional (RGB) components. This approach also allows the data to be downsampled to a lower resolution, preventing the model from being overly focused on fine details. Additionally, this method enhances the model's ability to intuitively identify patterns and anomalies in the data compared to traditional time-series representations. These advantages make PixleepFlow an explainable deep learning model that comprehensively captures the multifaceted nature of human behavior in multi-modal tasks.

PixleepFlow is designed to estimate sleep quality and stress level by converting synchronized sensor data into composite image data. The model evaluates these metrics using a total of seven indicators derived from daily survey records and sleep sensor data: overall sleep quality as perceived by the participant immediately after waking up (Q1), emotional state just before sleep (Q2), stress level experienced just before sleep (Q3), total sleep time (TST, S1), sleep efficiency (SE, S2), sleep onset latency (SOL, S3), and wake after sleep onset (WASO, S4). Moreover, to capitalize on the advantages of image translation, PixleepFlow incorporates Explainable Artificial Intelligence (XAI) techniques to provide visual interpretations of the results, enhancing the transparency and explainability of the model's predictions. The PixleepFlow framework is shown in Fig. \ref{fig:PixleepFlow}.

By utilizing PixleepFlow, we can more effectively assess both sleep quality and stress level while providing an intuitive, visual understanding of the data patterns and anomalies. This allows for a comprehensive analysis and deeper understanding of the complex factors associated with sleep.

Based on the predicting human lifelog metrics competition, we aim to discover the following:

1) Efficacious Methodology: Describe effective methodologies for achieving optimal performance in the competition.

2) Optimal Model Architectures: Identify which model architectures perform best on lifelog datasets by comparing deep learning approaches.

3) Data Format Suitability: Determine which data formats (raw signals, spectrograms, or images) most effectively represent lifelog datasets and yield the highest predictive F1 score.

4) Critical Channels: Analyze which specific sensor channels (e.g., acceleration, heart rate, Global Positioning System GPS) are most strongly correlated with stress level metrics such as sleep quality and emotional states. In this context, ``channels" refer to the distinct types of data streams collected by various sensors, each providing unique insights into physiological and environmental conditions.

By conducting this research, deriving these results using sensors that can be encountered in everyday life is thought to contribute significantly to human understanding in the future.

\renewcommand{\arraystretch}{1.5}
\setlength{\tabcolsep}{10pt}

\section{Methodology}
\label{Section:Methodology}

\begin{table}[h]
\centering
\caption{Overview of Smartphone and Smartwatch Sensor Data. The term 'ext.' refers to sensors used in extension experiments, while 'org.' refers to sensors used during the original model training.}
\label{tab:sensor_data}
\small
\resizebox{\columnwidth}{!}{%
\begin{tabular}{@{}l l c c c c c c@{}}
\toprule
\multirow{2}{*}{\textbf{Device}} & \multirow{2}{*}{\textbf{Sensor}} & \multirow{2}{*}{\textbf{Feature}} & \multicolumn{4}{c}{\textbf{Channels}} & \multirow{2}{*}{\textbf{Freq. (Hz)}} \\
\cmidrule(r){4-7}
 &  &  & \textbf{5} & \textbf{7} & \textbf{11} & \textbf{18} & \\
\midrule
\multirow{9}[+12]{*}{\centering\rotatebox[origin=c]{90}{\textbf{Smartphone}}}
 & \multirow{3}{*}{\makecell[l]{Accele-\\ration}} & x & \checkmark & \checkmark & \checkmark & \checkmark & \multirow{3}{*}{50} \\
\cmidrule{3-7}
 &  & y & \checkmark & \checkmark & \checkmark & \checkmark & \\
\cmidrule{3-7}
 &  & z & \checkmark & \checkmark & \checkmark & \checkmark & \\
\cmidrule{2-8}
 & Activity & activity & \checkmark & \checkmark & \checkmark & \checkmark & 1/60 \\
\cmidrule{2-8}
 & \multirow{4}{*}{\makecell[l]{GPS\\coord.}} & altitude &  &  & \checkmark & \checkmark & \multirow{4}{*}{1/5} \\
\cmidrule{3-7}
 &  & latitude &  &  & \checkmark & \checkmark & \\
\cmidrule{3-7}
 &  & longitude &  &  & \checkmark & \checkmark & \\
\cmidrule{3-7}
 &  & speed &  & \checkmark & \checkmark & \checkmark & \\
\cmidrule{2-8}
 & Ambient & ambience &  & \checkmark & ext. & \checkmark & 1/120 \\
\midrule
\multirow{9}[+12]{*}{\centering\rotatebox[origin=c]{90}{\textbf{Smartwatch}}} 
 & Light & m\_light &  &  & org. & \checkmark & 1/600 \\
\cmidrule{2-8}
 & \multirow{7}{*}{\makecell[l]{Step\\counts}} & burned\_cal. &  &  &  & \checkmark & \multirow{7}{*}{1/60} \\
\cmidrule{3-7}
 &  & steps &  &  & \checkmark & \checkmark & \\
\cmidrule{3-7}
 &  & distance &  &  &  & \checkmark & \\
\cmidrule{3-7}
 &  & running\_steps &  &  &  & \checkmark & \\
\cmidrule{3-7}
 &  & speed &  &  &  & \checkmark & \\
\cmidrule{3-7}
 &  & step\_freq. &  &  &  & \checkmark & \\
\cmidrule{3-7}
 &  & walking\_steps &  &  &  & \checkmark & \\
\cmidrule{2-8}
 & Heart rate & heart\_rate & \checkmark & \checkmark & \checkmark & \checkmark & 1/60 \\
\bottomrule
\end{tabular}
}
\end{table}

\subsection{Datasets}

The "Real-world Multimodal Lifelog Dataset" encompasses over 10,000 hours of data collected from Android smartphones and Empatica E4 smartwatch devices in 2020 \cite{chung2022real}.
Similarly, the 2023 data collection employed the same methodology as the 2020 study, with eight patients' data being gathered using Android smartphones and either the Galaxy Watch4 or Watch5 smartwatch device \cite{oh2024human}.
The 2023 collected dataset, comprising a total of 220 days from the lifelog data of 8 patients, is considered due to the differences between smartwatch devices. 
However, a significant limitation of this dataset is its relatively small size, with only 4 patients used for training and 4 patients for testing. The limited sample size increases the risk of overfitting, which restricts the generalizability of the findings and poses challenges in achieving robust model performance. 
The objective is to predict seven overall life quality labels (sleep quality, emotional state, stress strength) and sleep stats (total sleep time, sleep efficiency, SL, WASO) scores using F1 metrics. The detailed evaluation metrics are described by Oh et al \cite{oh2024human}.

\subsection{Analysis of Selected Sensor Signals}
\label{Section II, B}
5 or 11 of the 18 signals in the dataset were employed for model training, respectively. The 5 channels composite image incorporated the three-axis acceleration, activity, and heart rate data obtained from the smartphone and smartwatch, while the 11 channels composite image augmented the dataset with GPS coordinates, light intensity, and step count information.
Detailed signal information is described in TABLE \ref{tab:sensor_data}. We selectively utilized these signals for training based on the following criteria. (1) Three-axis acceleration: Measured continuously for 24 hours provides a rough estimation of movement information during the day and night, and is often employed in sleep studies \cite{yoshihi2021estimating}. The acceleration signal measured by the smartphone has a frequency of 50 Hz, which allows for the indirect determination of the sleep onset time and wake-up time.
(2) Activity: Behavioral classification is recorded once per minute (1/60 Hz), indicating whether the user is in motion. This classification can be combined with acceleration data to enhance the accuracy of activity detection.
(3) GPS Coordinates: These include altitude, latitude, longitude, and speed values, collected at a frequency of 1/5 Hz, which facilitates the analysis of the movement and bedtime location of users. (4) Light intensity: The state of the lights can be used to infer whether the user is sleeping. (5) Steps: Reflects the activity level, potentially impacting the total sleep time. (6) Heart rate: Indicates various body states, which decrease during sleep compared to wakefulness, and it is well-established that severe anxiety and stress cause an increase in heart rate.

Signals collected in real-life environments, where factors like room brightness and noise are not well controlled, tend to be noisy, leading to lower prediction accuracy. Therefore, to obtain more accurate results, we selectively used only signals highly related to sleep or stress.

\begin{figure}[ht]
\centering
\includegraphics[width=0.5\textwidth]{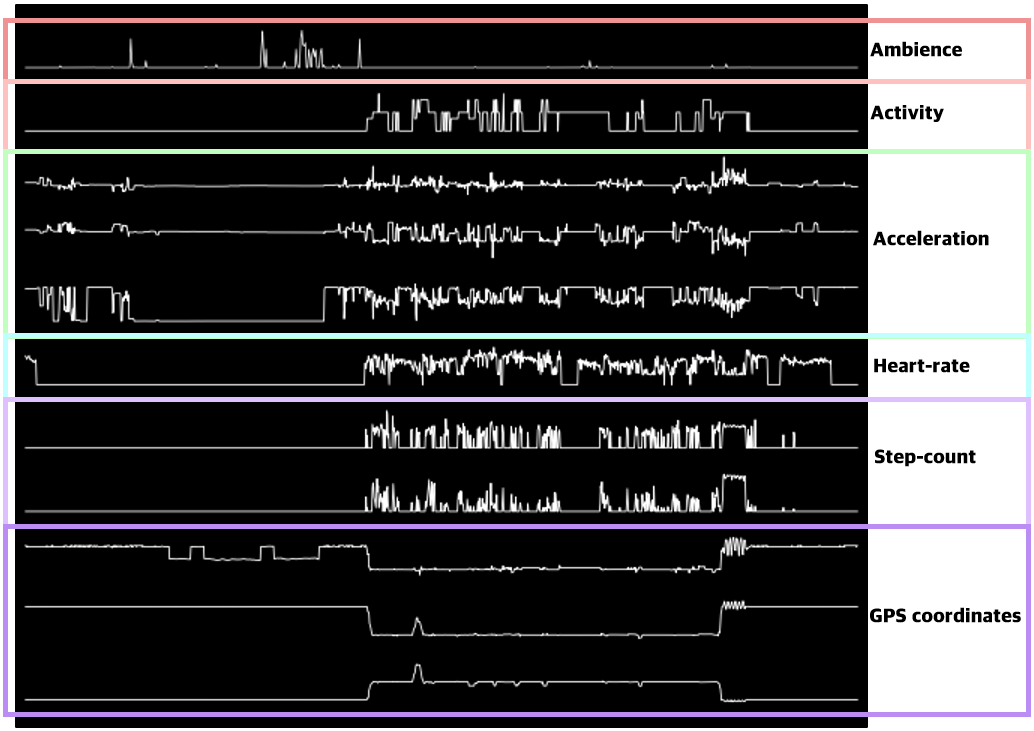}
\caption{Sample image of the synchronized 11 channels dataset mentioned in TABLE \ref{tab:sensor_data}.
The x-axis consists of 86,400 data points, representing 1-second intervals, showing synchronized data.
}
\label{fig:data}
\end{figure}

\subsection{Data Pre-processing}
\label{Section:Data pre-processing}
It should be noted that the provided dataset does not contain data for all hours of the day. Consequently, there may be instances when data is absent for a given duration, which varies by sensor. To illustrate, the heart rate channel is unavailable from midnight until morning in Fig.\ref{fig:data}. In addition, the frequency of data collection (Hz) differs from sensor to sensor, as well as the time of day when the data was collected. Therefore, a synchronization process is required to ensure the effective utilization of multiple sensors. Our proposed PixleepFlow approach to this issue is as follows:

\subsubsection{Frequency resampling}
Due to the varying collection frequencies of different sensors, it was essential to normalize these frequencies to achieve synchronization. We opted to aggregate data on a per-second basis (1Hz), aligning the nearest data point to each second.

\subsubsection{Signal Synchronization}
The data were then synchronized into a single dataset spanning from 00:00:00 to 23:59:59, representing a total of 86,400 seconds (24 hours). Every second was recorded, and the synchronized data can be transformed into various data types. 

\subsubsection{Missing values interpolation} 
\label{subsection2:missing_values}
For data with missing values, we needed a method of interpolation. We used linear interpolation. Moreover, we leave the data as it is without interpolating where there is no data at the beginning and end of the data. This is to avoid generating too much spurious data.

\subsection{Various Transformations of Signal Data}
% As previously stated in Section \ref{Section:Methodology}, an inherent drawback is the need for interpolation when configuring the matrix to integrate various types of raw data.
Following the preprocessing methodologies stated in Section \ref{Section:Methodology}, the following approaches illustrate different ways to transform and utilize the signal data:

\subsubsection{Using Raw signal data}
The method of leveraging the raw signal data is to utilize the value vectors captured by the sensor in their original form. By inputting each time series dataset into the model in its unprocessed state and conducting analysis, it is possible to preserve the integrity of the original data to the greatest extent. This approach allows for the direct utilization of raw data, facilitating the discovery of correlations and temporal changes between sensors without requiring additional processing.

\subsubsection{Using Spectrogram data}
A spectrogram is a visual representation of the frequency changes of a signal over time, displaying the signal's spectrum as it varies along the time axis. This transformation permits a more precise examination of the data's characteristics by transforming each sensor's data into a spectrogram, which captures both the temporal and frequency characteristics, relying on the raw time-series data. Similar to \cite{neshitov2021wavelet}, we employed the Short-Time Fourier Transform (STFT) for this conversion.

\subsubsection{Using Imaging Data}
\begin{figure}[ht]
\centering
\begin{minipage}{0.22\textwidth}
    \centering
    \includegraphics[width=\textwidth]{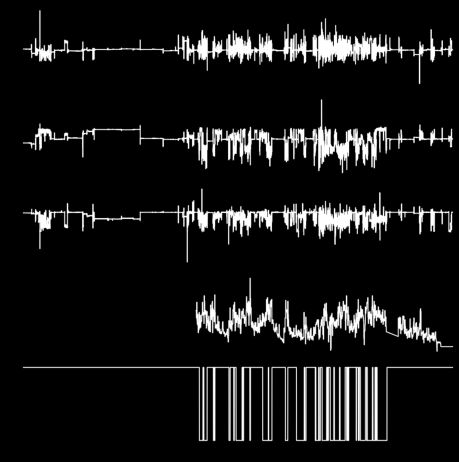}
    \caption*{(a) 5-channel Image}
\end{minipage}\hfill
\begin{minipage}{0.22\textwidth}
    \centering
    \includegraphics[width=\textwidth]{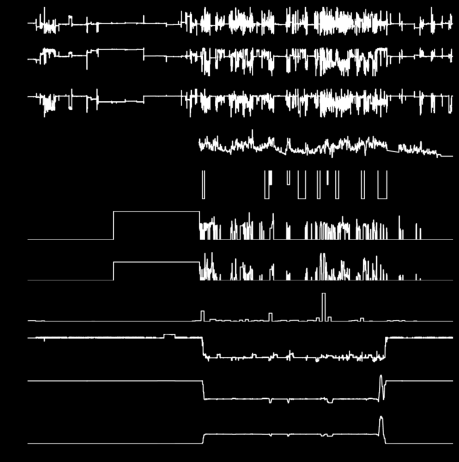}
    \caption*{(b) 11-channel Image}
\end{minipage}\hfill
\caption{Data Visualization of 5 and 11 channels}
\label{fig:visualization_channel}
\end{figure}

A methodology was employed to visualize raw time-series data by representing sensor values over time in a single image. Converting data into images adheres to the methodology delineated in Section \ref{Section:Data pre-processing}. Using the subplot function in Matplotlib, the visualizations for each sensor were stacked vertically. This approach comprehensively captures the temporal variations of each sensor, and image processing techniques were applied to these visual representations for model training.
The data were transformed into diverse image formats using both 5 channels and 11 channels data, with the 5 channels image data demonstrating the most optimal performance. This approach presents a novel method for effectively analyzing and processing complex time-series data, with the transformed data depicted in Fig \ref{fig:visualization_channel}.

\section{EXPERIMENTS}

\subsection{Experimental Setup}
To optimize the model, we used the AdamW optimizer and set the weight decay to 0.1. To build a model that is robust to different user data distributions, we used the K-fold ensemble technique with random sampling, with 200 epochs of training for each fold. For each fold, we selected the model with the highest score, and the final score was calculated by averaging these best scores across all folds. The macro F1-score was used as the model storage criterion. Each class has a binary label, and Binary Cross Entropy Loss was used. The initial learning rate was set to 5e-5, and the cosine annealing scheduler was set to T=200 to progressively decrease the learning rate throughout training. All implementations were done using the PyTorch library, and training was performed on an NVIDIA H100 GPU.

\subsection{Experimental Approach}
We explored the efficacy of various view experiments by converting data in diverse formats, extending beyond the conventional approach of simple image conversion techniques. To assess the model's performance, we utilized a range of data types, including raw sensor data, composite images with various channels, and spectrograms which are mentioned in Section \ref{Section:Methodology}.

\subsubsection{Raw-based}
In the raw signal data format, the signal is embedded using 1D-CNN layers with the ResNet mechanism applied. Multiple synchronized channels of sensor values, which have been transformed as described in Section \ref{Section:Data pre-processing}, are inputted in their original form, allowing the model to learn directly from the intuitive data without any additional transformations. This approach not only preserves the characteristics of the original data but also maintains temporal continuity, enabling the model to effectively capture time-series patterns. Additionally, by handling 1D signals, the model learns clear patterns such as specific frequency components or temporal sequences, all while utilizing a relatively simple model structure that requires less computational resources.

\subsubsection{Image-based (PixleepFlow)}
For PixleepFlow, we used the SEResNeXt101\_32x4d and ResNeXt101\_32x32d \cite{xie2017aggregated,hu2018squeeze} models, which were pre-trained on the ImageNet 21k dataset. 

ResNeXt is a highly modularized network architecture that employs a cardinality dimension, allowing for increased model capacity and diversity without significantly increasing computational complexity. It builds upon the ResNet architecture by utilizing grouped convolutions, which enable multiple transformations to be processed in parallel, leading to enhanced feature learning capabilities \cite{xie2017aggregated}.

SE-ResNeXt extends the ResNeXt architecture by integrating Squeeze-and-Excitation (SE) blocks, which adaptively recalibrate channel-wise feature responses. These SE blocks enhance the network's sensitivity to informative features by explicitly modeling the interdependencies between channels \cite{hu2018squeeze}. The combination of ResNeXt's cardinality and SE blocks results in a powerful and efficient architecture capable of capturing complex patterns in the data.

\subsubsection{Spectrogram-based}
Each sensor data converted into a spectrogram is individually extracted while passing through ResNet-based 1D-CNN layers. Features extracted from each sensor channel are combined into one integrated feature vector through the concatenate process. These combined feature vectors are input to the final classifier model to execute classification. Through this process, more accurate behavioral pattern recognition is possible by using temporal-frequency information of each sensor data.

\begin{figure}[ht]
\centering
\begin{minipage}{0.155\textwidth}
    \centering
    \includegraphics[width=\textwidth]{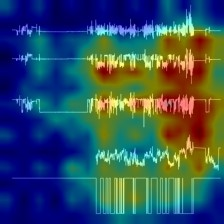}
\end{minipage}\hfill
\begin{minipage}{0.155\textwidth}
    \centering
    \includegraphics[width=\textwidth]{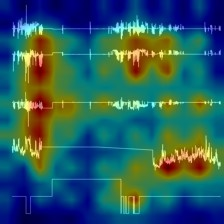}
\end{minipage}\hfill
\begin{minipage}{0.155\textwidth}
    \centering
    \includegraphics[width=\textwidth]{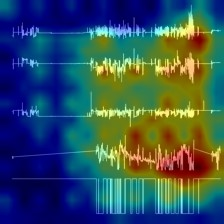}
\end{minipage}
\caption{Full-CAM visualization of XAI (Explainable Artificial Intelligence) highlighting the features utilized for sleep-related activity detection.} 
\label{fig:xai}
\end{figure}

\subsection{Model Interpretation}
PixleepFlow employs Explainable Artificial Intelligence (XAI) methodologies to achieve optimal performance and elucidate the rationale behind the model's predictions in a manner that is intelligible to humans. XAI is a technique that renders the output and decision-making process of machine learning models transparent, thereby facilitating user comprehension and trust in the model's behavior. In this study, we employed the Full-CAM (Full-Gradient Representation for Neural Network Visualization) technique \cite{srinivas2019full}. Full-CAM offers a visual representation of the impact of each input image on a specific output class within the network, emphasizing the image regions essential for the model to make certain decisions. The results obtained via this technique are illustrated in Fig \ref{fig:xai}.

As illustrated in Fig. \ref{fig:xai} the Full-CAM reveals that the activity information is not a significant indicator. Conversely, accelerometer (ACC) channels, particularly the Y and Z axis sensors and heart rate were frequently utilized as pivotal features \cite{yoshihi2021estimating, radha2019sleep}. Additionally, to identify features related to sleep, the primary prediction category in this competition, models demonstrated a tendency to focus on movements occurring immediately before or following approximate time in bed (TIB). Even when the participant checked their smartphone during the sleep period, the models showed a tendency to prioritize these subtle movements, even though they were less pronounced compared to the noticeable movements observed during wakefulness as shown in the central image Fig. \ref{fig:xai}.

\subsection{Result}
\begin{table}[h]
\centering
\caption{Model Performance Based on Channel and Input Type}
\label{tab:model_performance}
\small
\resizebox{\columnwidth}{!}{
\begin{tabular}{@{}l c c c@{}}
\toprule
\multirow{2}{*}{\textbf{Channel}} & \multicolumn{3}{c}{\textbf{Input Type}} \\
\cmidrule(r){2-4}
 & \textbf{Image-based (pixleep)} & \textbf{Raw-based} & \textbf{Spectrogram-based} \\
\midrule
\textbf{5} & 0.680 & 0.636 & 0.613 \\
\textbf{11} & 0.746 & 0.699 & 0.637 \\
\bottomrule
\end{tabular}
}
\end{table}

\begin{table}[h]
\centering
\caption{Model Performance Based on Channel and Input Type. The meaning of labels Q1 through S4 can be referenced in Fig \ref{fig:PixleepFlow}.}
\label{tab:pixleep_metrics}
\small
\resizebox{\columnwidth}{!}{%
\begin{tabular}{@{}lccccccc@{}}
\toprule
\multirow{2}{*}{\textbf{Channel}} & \multicolumn{7}{c}{\textbf{Image-based (Pixleep)}} \\ 
\cmidrule(r){2-8}
 & \textbf{Q1} & \textbf{Q2} & \textbf{Q3} & \textbf{S1} & \textbf{S2} & \textbf{S3} & \textbf{S4} \\ 
\midrule
\textbf{5}  & 0.762 & 0.800 & 0.308 & 0.615 & 0.700 & 0.857 & 0.714 \\ 
\textbf{11} & 0.706 & 0.804 & 0.767 & 0.593 & 0.751 & 0.855 & 0.749 \\ 
\bottomrule
\end{tabular}%
}
\end{table}

In this study, we undertake a comparative analysis of the performance of data from 5 and 11 channels for the Image (PixleepFlow), raw, and spectrogram modalities. The results of the experiment are presented in TABLE \ref{tab:model_performance}. The image-based  (PixleepFlow) demonstrated the most optimal performance in both channel configurations, attaining the highest F1-score of 0.746 on 11 channels of data, which is notably higher than 0.680 on 5 channels of data. The raw-based method yielded an F1-score of 0.636 for the 5 channels model and 0.699 for the 11 channels model. In comparison, the spectrogram-based approach exhibited the lowest performance of the three methods, with an F1-score of 0.613 for the 5 channels model and 0.637 for the 11 channels model. 
These results indicate that our proposed PixleepFlow approach not only achieves superior performance with 11 channels but also demonstrates comparable effectiveness with just 5 channels, achieving a performance level close to that of the raw-based method with 11 channels.

As seen in TABLE \ref{tab:pixleep_metrics}, the improvement in stress level (Q3) prediction observed in the 11 channels configuration, is likely due to the additional sensor data providing more informative cues. While the F1-score for Q3 was very low in the 5 channels setup (0.308), it significantly improved in the 11 channels setup (0.767). This suggests that diverse sensor data play a crucial role in accurately predicting stress level. However, the F1-score for S1 (sleep duration) remains low (11 channels: 0.593), indicating that sleep duration is a complex feature that may be challenging to capture fully with single image data or individual sensor inputs.

Thus, PixleepFlow demonstrates strong performance in predicting certain labels, such as emotional state (Q2: 0.804) and sleep onset latency (S3: 0.855), but the prediction of stress level (Q3: 0.767) and sleep duration (S1: 0.593) shows significant variance depending on the channel configuration. This indicates that integrating various sensor data could potentially enhance predictive performance.

Overall, as the number of channels increased, the performance tended to improve across all input types. It is interpreted that more channels contributed to increasing the accuracy of the classification task by providing richer information to the model. 

\section{Extended Comparative Analysis}
\label{Section:Extended Comparative Analysis}
In this experiment, we conducted independent tests separate from the competition to achieve more precise results. We examined the differences in performance based on various optimizers, schedulers, and the number of channels (5, 7, 11, and 18 channels, as seen in TABLE \ref{tab:sensor_data}), as well as the impact of different frequencies (1 Hz and 1/60 Hz).

\subsection{Extended Experimental Setup}
The experimental setting is analogous to the existing model, yet it exhibits several discernible distinctions. The model is fixed with SERexNeXt101\_32x4d and uses a model pre-train with the ImageNet 21k dataset. Furthermore, if the loss was 0.8 or higher in epoch 100 or beyond, early stopping was conducted beyond the reference value, which may indicate an increased risk of overfitting.

\subsubsection{SAM optimizer}
One method for mitigating overfitting due to the limited dataset is to utilize a SAM optimizer \cite{foret2020sharpness}. The SAM optimizer, which stands for Sharpness-Aware Minimization, has been demonstrated to enhance a model's generalization performance. By flattening the surface of the model's loss function, SAM optimizers facilitate improved generalization to novel data, reducing the probability of overfitting to a specific dataset.

\subsubsection{SGDR Scheduler}
Another generalization method employed is the Stochastic Gradient Descent with Warm Restarts (SGDR) scheduler. The SGDR method involves the periodic resetting of the learning rate, to achieve enhanced optimization. 

\subsubsection{Add Ambience Data}
For channels 7, 11, and 18, we incorporated the most recent ambient sound data. This dataset includes the top 10 probability values for various sounds detected by the smartphone, such as "breathing," "snoring," and "snorting," all of which are closely associated with sleep. Ambient sound data typically showed higher values during sleep periods, which may enhance the model's ability to capture relevant sleep-related information, as demonstrated in Fig. \ref{fig:data}. We hypothesize that integrating ambient sound data can lead to more accurate predictions of sleep efficiency. For a comprehensive breakdown of the utilized sensor data, please refer to TABLE \ref{tab:sensor_data}.

\subsubsection{Frequency test}
In the preceding experiment, a total of 86,400 data points were recorded at a rate of 1 Hz. To confirm the impact of Hz on performance, an additional experiment was conducted, in which the data were converted to units per minute (1/60 Hz). While this may result in the aggregation of subtle differences in the data, it may also mitigate the phenomenon of the model overfitting to fine features.

\subsection{Experiment}
\begin{table}[h]
\centering
\caption{Model Performance Based on Channel and Frequency}
\label{tab:extend_result}
\small
\resizebox{\columnwidth}{!}{%
\begin{tabular}{@{}l l c c c c c c@{}}
\toprule
\multirow{2}{*}{\textbf{Metric}} & \multirow{2}{*}{\textbf{Frequency}} & \multicolumn{4}{c}{\textbf{Channels}} \\
\cmidrule(r){3-6}
 &  & \textbf{5} & \textbf{7} & \textbf{11} & \textbf{18} \\
\midrule
\multirow{2}{*}{\textbf{Image-based (Pixleep)}} & \textbf{1/60Hz} & 0.612 & 0.629 & 0.635 & 0.638 \\
\cmidrule(r){2-6}
 & \textbf{1Hz} & 0.659 & 0.521 & 0.634 & 0.632 \\
\bottomrule
\end{tabular}
}
\end{table}

The mean score of our K-fold (k=5) indicated that the 5 channels data at 1 Hz yielded the optimal result, with a score of 0.659. This is the same sensor utilized in the initial submission. Subsequently, the 11, 18, and 7 channels at 1 Hz yielded scores of 0.634, 0.632, and 0.521, respectively. Additionally, the 18 channels, 11 channels, 7 channels, and 5 channels at 1/60 Hz yielded scores of 0.638, 0.635, 0.629, and 0.612, respectively. The experimental results are summarized in TABLE \ref{tab:extend_result}.

In general, the most stable frequency was 1/60 Hz, with the highest value being the 5 channels data at 1 Hz. In light of the interchannel interactions, it is probable that the 5 channels dataset comprised solely the indispensable sensor information and exhibited optimal interaction. The higher frequency of 1 Hz provides more data, enabling fine-grained pattern learning; however, including an excessive number of channels can impede the model's capacity for generalization. The data collected at 1/60 Hz was less complex, allowing the model to learn with greater reliability, particularly in the case of the 18 channels data, which may be attributed to the effective integration of information from all channels. Additionally, certain sensors, such as the accelerometer and heart rate, played a pivotal role, and the 5 channels data may have demonstrated superior performance due to its effective inclusion of information from these crucial sensors \cite{radha2019sleep}.

\section{Conclusion}
In conclusion, this study presented a comprehensive approach to predicting sleep quality and stress level through the innovative Pixel Sleep Flow Framework (PixleepFlow). By converting multimodal lifelog sensor data into composite images, PixleepFlow enhances the analysis of sleep patterns and their impacts on health. This methodology not only improves the accuracy of predicting key metrics such as total sleep time, sleep efficiency, and stress level but also provides a visually interpretable model by leveraging Explainable Artificial Intelligence.

The findings underscore the significance of synchronization and transformation of sensor data, highlighting the efficacy of the image conversion approach in capturing the multifaceted nature of human behavior. The results from the Full-CAM visualization confirmed the importance of specific sensor channels, such as accelerometer and heart rate, in predicting sleep and stress-related metrics.
Additionally, by leveraging the XAI technique, we were able to mitigate concerns of overfitting due to the limited dataset size and ensure that the model was learning effectively from the available data. 

Future research can build upon this work by exploring additional sensor combinations and refining the image conversion techniques to improve prediction accuracy. The application of PixleepFlow in real-world settings holds the potential for personalized health monitoring and promoting overall well-being through more accurate and interpretable assessments of sleep and stress.

\printbibliography

% \bibliographystyle{IEEEtran}  % IEEEtran 스타일 사용
% \nocite{*}  % 강제로 모든 참고문헌 출력

% \addbibresource{references.bib} % Add the references.bib file
% \bibliography{conference_101719}  % .bbl 없이 .bib 파일명을 입력
%\bibliography{conference_101719}  % .bbl 없이 .bib 파일명만 입력

\vspace{12pt}

\end{document}